\documentclass[pra,twocolumn,titlepage,nofootinbib,amsmath,showpacs]{revtex4}%
\usepackage{graphicx}%
\begin{document}
\title{The recoil correction to the proton-finite-size contribution to the Lamb shift in muonic hydrogen}
\author{Savely~G.~Karshenboim}
\email{savely.karshenboim@mpq.mpg.de}
\affiliation{Max-Planck-Institut f\"ur Quantenoptik, Garching,
85748, Germany} \affiliation{Pulkovo Observatory, St.Petersburg,
196140, Russia}
\author{Evgeny Yu. Korzinin}
\affiliation{D.~I. Mendeleev Institute for Metrology, St.Petersburg,
190005, Russia}
\author{Vladimir G. Ivanov}
\affiliation{Pulkovo Observatory, St.Petersburg, 196140, Russia}
\author{Valery A. Shelyuto}
\affiliation{D.~I. Mendeleev Institute for Metrology, St.Petersburg,
190005, Russia}

%\date

%\today

\begin{abstract}
The Lamb shift in muonic hydrogen was measured some time ago to a
high accuracy. The theoretical prediction of this value is very
sensitive to the proton-finite-size effects. The proton radius
extracted from muonic hydrogen is in contradiction with the results
extracted from elastic electron-proton scattering.
That creates a certain problem for the interpretation of the results
from the muonic hydrogen Lamb shift. For the latter we need also to
take into account the two-photon-exchange contribution with the proton
finite size
involved. The only way to describe it relies on the data from the
scattering, which may produce an internal inconsistency of theory.

Recently the leading proton-finite-size contribution to the
two-photon exchange was found within the external field
approximation. The recoil part of the two-photon-exchange has not
been considered. Here we revisit calculation of the external-field
part and take the recoil correction to the finite-size effects into account.
\end{abstract}
\pacs{
{12.20.-m}, %Quantum electrodynamics
{13.40.Gp}, %Electromagnetic form factors
{31.30.J-}, %Relativistic and quantum electrodynamic effects in atoms and molecules
{36.10.Ee}, %Muonium, muonic atoms and molecules
% {32.10.Fn} %Fine and hyperfine structure
% {13.60.Fz} %Elastic and Compton scattering
%
}
\maketitle

%\tableofcontents

\section{Introduction}

A few years ago the Lamb shift ($2p_{1/2}-2s_{1/2}$) in muonic hydrogen
was experimentally determined to a great accuracy \cite{Science}. The related
theoretical prediction is much less accurate, suffering from a bad
knowledge of the proton charge radius. Therefore, the experimental data and theoretical expression have been utilized to
determine an accurate value of the proton radius. Actually, theory
(see, e.g., \cite{VASH-book,Science,Jentschura2011}) disagrees with
 experiment and therefore the muonic-hydrogen value of the proton
charge radius is inconsistent with results of hydrogen spectroscopy
\cite{codata2010} and evaluations of the elastic electron-proton
scattering (see, e.g., \cite{mami,sick,zhan,sick2014}).

The theoretical prediction for the Lamb shift in muonic hydrogen is basically of the form
\begin{equation}\label{Eq:shape}
\Delta E_L = c_1 + c_2 \, R_E^2 \;,
\end{equation}
where the first term, dominated by the Uehling correction and other
QED effects, does not depend on the proton rms electric charge radius
$R_E$, and the second term is dominated by the leading
proton-finite-size correction. Meanwhile, there is a higher-order
proton-finite-size correction, which does depend on the distribution
of the proton charge (or rather we prefer to discuss the
electric charge form factor of the proton $G_E(q^2)$), but does not have the
required form of ${const} \times R_E^2$.

The general problem in its calculation is that to find it one has to
use a certain description of the charge distribution (or
$G_E(q^2)$). There is no satisfactory {\em ab initio\/} model. So,
one either could apply a certain variety of models, which are not
really related to experimental data and arbitrary to a certain
extent, and to estimate somehow the uncertainty, or, alternatively,
one could take advantage of the existing experimental data. The
first option does not really work for the proton. As for the second
option, the experimental data by themselves are not accurate enough
and one cannot use the data directly, Instead one has to apply a
fit. Unfortunately, and that is a real problem with the proton
radius, the fits also produce their value of $R_E$ and those values
from empiric fits do not agree with the value from the muonic
hydrogen (see, e.g., parameters of low momentum expansion of some
empiric fits in Table~\ref{t:exp:fits}).

We have to clarify that
the correction we discuss here is a small one, while the
contradiction is huge. The very fact and the scale of the
discrepancy does not depend on the way we treat the higher-order
proton-finite-size term. Meanwhile, the eventual value of the muonic $R_E$
does depend on the way, we treat the higher-order finite-size correction,
at the level of the uncertainty of $R_E$. Here we revisit the
correction.

The higher-order proton-finite-size correction can be understood as
a certain moment of the charge distribution. Since the very mean
radius of that distribution is under question because the results
from the scattering and the muonic hydrogen are different, we cannot
directly apply the scattering results for these corrections
(see, e.g., \cite{pla,Fri:mami,my_adp,efit}).

Thus, the next-to-leading proton-finite size corrections are
sensitive to detail of the charge distribution. A self-consistent
consideration of those shape-sensitive contributions was performed
in \cite{efit}. The consideration was done within the external field
approximation, which delivers the leading part of the
shape-sensitive contributions. Here, we present a complete
evaluation, including recoil effects. Those involve not only the charge
distribution, but also a distribution of the magnetic moment.

There are two main ingredients of our consideration. We have to explain the method of a treatment of the higher-order finite-size correction within a theory of muonic hydrogen, which would be consistent both with the experimental data on the electron-proton scattering and with the muonic value of the proton radius. The method has been applied previously \cite{efit} within the external field approximation. Here we are to expand it to accommodate the recoil finite-size corrections, which we also need to consider. For that we follow \cite{carlson,birse}. In both cases we start following cited papers, but modify the approaches.

\section{General expression for the elastic two-photon-exchange contribution}

\begin{figure}[htbp]
\begin{center}
\resizebox{0.2\columnwidth}{!}{\includegraphics{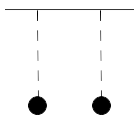}}
\end{center}
\caption{The diagram for the Friar correction to Lamb shift in muonic hydrogen. The appropriate subtractions are assumed. \label{Fig:friar}}
\end{figure}

Let's start with the expression for the proton-finite-size contribution including the recoil effects. They are described by the two-photon exchange (TPE). The leading part of the higher-order proton-finite-size term is the so-called Friar
term. That is a result derived within the external field approximation
(see Fig.~\ref{Fig:friar}). The result for the correction was first derived in the
coordinate space \cite{friar79s,friar79,bortr} and next in the momentum space \cite{pachucki96muon} (see \cite{VASH-book} for details of the derivation)
\begin{eqnarray}\label{Eq:friar:1}
\Delta E_{\rm Fr}(nl) &=& - \frac{16(Z\alpha)^5\,m_r^4}{\pi}\, I_{\rm Fr}\frac{\delta_{l0}}{n^3}\;,\\
I_{\rm Fr}&=&\int_0^\infty
{\frac{dq}{q^4}}\left[\left(G_E(q^2)\right)^2-1-2G_E^\prime(0)\,q^2\right]\;.\nonumber
\end{eqnarray}
Here and throughout the paper we apply the relativistic units, in which $\hbar=c=1$, $m$ stands for the muon mass, $M$ is  the proton mass and $m_r$ is the reduced mass. The momentum $q$ is the Euclidean one.

\begin{figure}[htbp]
\begin{center}
\resizebox{0.5\columnwidth}{!}{\includegraphics{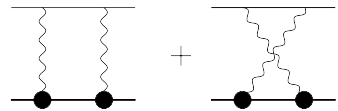}}
\end{center}
\caption{Two-photon-exchange: elastic contribution. Necessary subtractions, which follow from an EDE consideration, are assumed. The circles are for the vertex with the appropriate form factors.\label{Fig:elastic}}
\end{figure}

\begin{figure}[htbp]
\begin{center}
\resizebox{0.5\columnwidth}{!}{\includegraphics{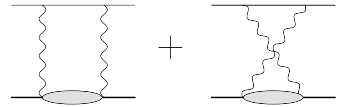}}
\end{center}
\caption{Two-photon-exchange: inelastic contribution (the proton polarizability).\label{Fig:polar}}
\end{figure}

If we like to consider the TPE correction with all recoil effects included, we have to use a certain two-body formalism, such as the effective Dirac equation (EDE) \cite{ede,SYQED,VASH-book}. Within such an approach the TPE diagrams appear with certain subtractions (related to proton pole contributions) and on top of those subtracted two-body TPE diagrams one has to add a perturbative series of an effective one-body theory, e.g., the EDE theory with a proton on mass shell. The latter is presented by the Friar contribution, while the former is related to two types of the effects. Those include elastic (with subtractions) and inelastic (polarizability) parts (see Figs.~\ref{Fig:elastic} and~\ref{Fig:polar}), which at a certain stage are to be considered together.

Indeed, once we have well-defined elastic and inelastic quantities, we can work on their calculation independently. However, their definition is not a straightforward issue. As an example one can consider a calculation of the elastic part. It requires an appropriate on-shell vertex for the proton. One may deal with the Dirac and Pauli form factors $F_{1,2}$ or with the Sachs ones $G_{E,M}$. The difference between those two
parametrizations leads to a contribution, which vanishes for the
on-shell vertex, but is not equal to zero for the TPE diagrams in Fig.~\ref{Fig:elastic}, which involve the vertexes off-shell. In other words, the elastic part is not well-defined by itself.

Meanwhile, both elastic and inelastic contributions have their
specifics and their calculation is very different.  Indeed, there are distinguished contributions into each of them and
those contributions can be studied independently of the definitions.
One of them is the Friar term, the dominant part of
the elastic TPE and we indeed can study it independently of anything
else. (As we mention, the ambiguity in a possible definition
of the elastic term concerns usually only recoil effects, while the Friar
term is the result within the external field approximation.) Indeed,
the actual expression for the elastic part for the
two-photon-exchange (eTPE) correction is more complicated than the
Friar contribution (\ref{Eq:friar:1}). Once we intend to complete
the calculation of the elastic part and to add recoil corrections to
the Friar term, we have to check the definitions.

Therefore, we
have to explicitly quote the definitions of the elastic and inelastic TPE and to be sure they are
consistent. While calculations of the elastic and inelastic parts of TPE have been done in a single paper for a few occasions \cite{Pachucki1999,martynenko1999,martynenko2006}, the first complete evaluation of TPE, where consistency of their definitions was discussed, appeared some time later in \cite{carlson}. Unfortunately, it did not manage to produce consistent definitions of the
elastic and inelastic part. The consistency was achieved later in \cite{birse}.

The elastic part of the TPE contribution consists of a QED result for a point-like proton and of a proton-structure correction. We acknowledge that the point-like part has been already calculated and it is subtracted while calculating the eTPE term. The QED TPE term is the well-known Salpeter contribution (see, e.g., \cite{SYQED,VASH-book}). It is for the proton without any anomalous magnetic moment, which satisfies the conditions $F_1(0)=1, F_2(0)=0$ or $G_E(0)=G_M(0)=1$.

The analytic expression for the eTPE contribution was considered in \cite{Pachucki1999,martynenko1999,carlson,birse}. Here, we re-arrange the expression for our purpose. The general structure is of the form \cite{carlson,birse} (cf. \cite{Pachucki1999,martynenko1999})
\begin{eqnarray}\label{Eq:ETPE}
\Delta E_{\rm eTPE}(nl) &=& - \frac{16(Z\alpha)^5\,m_r^4}{\pi}\, I_{\rm eTPE}\frac{\delta_{l0}}{n^3}\;,\nonumber\\
I_{\rm eTPE}&=&I_{\rm Fr}+I_{\rm rec}\;,\nonumber\\
I_{\rm Fr}&=&\int_0^\infty{\frac{dq}{q^4}}\left[\left(G_E(q^2)\right)^2-1-2G_E^\prime(0)\,q^2\right]\;,\nonumber\\
I_{\rm rec}&=&I_{\rm E}+I_{\rm M}+I_{\rm F}\;,\nonumber\\
I_{\rm E}&=&\int_0^\infty{\frac{dq}{q^4}}f_E(m,M;q^2)\left[\left(G_E(q^2)\right)^2-1\right]\;,\nonumber\\
I_{\rm M}&=&\int_0^\infty{\frac{dq}{q^4}}f_M(m,M;q^2)\left[\left(G_M(q^2)\right)^2-1\right]\;,\nonumber\\
I_{\rm F}&=&\int_0^\infty{\frac{dq}{q^4}}f_F(m,M;q^2)\left[\left(F_1(q^2)\right)^2-1\right]\;.
\end{eqnarray}
The Dirac form factor
\[
F_1(q^2)=\frac{G_E(q^2)+\frac{q^2}{4m_p^2}\,G_M(q^2)}{1+\frac{q^2}{4m_p^2}}
\]
is more fundamental than the Sachs electric and magnetic form
factors $G_{E}$ and $G_M$. However, just the latter are
used to be fitted in empiric-fit papers.

The last term, $I_{\rm F}$, was excluded in \cite{carlson} from the
elastic part and attributed to the polarizability. Later on, it was
re-established by \cite{birse} as a part of the elastic TPE. Most of
the existing results on the polarizability
\cite{Pachucki1999,martynenko2006,birse,nevado,alarcon,peset} are
consistent with this definition\footnote{Actually, the very
calculation of the polarizability in \cite{carlson} is consistent
with the definition of the elastic part including this term as was
shown in \cite{birse}.} or,  as in case of \cite{gorchtein}, can be
re-adjusted to it.

The first term apparently is the Friar term, which is a non-recoil
contribution. The other terms are for the recoil. As it happens in
the two-body systems (see, e.g., various corrections to the Lamb
shift in \cite{VASH-book}) the external-field contributions and the
recoil one are treated differently and they have different
subtractions. The structure $G^2-1$ is a result of the subtraction of the expression for the point-like proton.

The explicit presentation of the $f$ factors \cite{carlson} is
\begin{eqnarray}\label{Eq:tpe:f}
f_E(m,M;q^2)&=&\frac{q}{2(M-m)}\left[\left(\frac{\gamma_2(\tau_p)}{\sqrt{\tau_p}}-\frac{\gamma_2(\tau_\mu)}{\sqrt{\tau_\mu}}\right)\frac{1}{1+\tau_p}
\right.\nonumber\\ &&\left.
-\left(\frac{1}{\sqrt{\tau_p}}-\frac{1}{\sqrt{\tau_\mu}}\right)\right]\;,\nonumber
\\
f_M(m,M;q^2)&=&\frac{q\,\tau_p}{2(M-m)}\left[\left(\frac{\gamma_2(\tau_p)}{\sqrt{\tau_p}}-\frac{\gamma_2(\tau_\mu)}{\sqrt{\tau_\mu}}\right)\frac{1}{1+\tau_p}
\right.\nonumber\\ &&\left.
-\left(\frac{\gamma_1(\tau_p)}{\sqrt{\tau_p}}-\frac{\gamma_1(\tau_\mu)}{\sqrt{\tau_\mu}}\right)\right]\;,
\nonumber \\
f_F(m,M;q^2)&=&\frac{M+m}{m}\,\tau_p\gamma_1(\tau_\mu)
\;,
\end{eqnarray}
where
\begin{eqnarray}
\tau_p&=&\frac{q^2}{4M^2}\,,\nonumber\\
\tau_\mu&=&\frac{q^2}{4m^2}\;,\nonumber\\
\gamma_1(\tau)&=&(1-2\tau)\left(\sqrt{1+\tau}-\sqrt{\tau}\right)+\sqrt{\tau}\;,\nonumber\\
\gamma_2(\tau)&=&(1+\tau)^{3/2}-\tau^{3/2}-\frac{3}{2}\sqrt{\tau}\;.\label{Eq:tau:p}
\end{eqnarray}

We are going to treat electric and magnetic form factors somewhat different and we are to decompose the Dirac form factor
\begin{eqnarray}
F_1^2-1&=&\frac{1}{(1+\tau_p)^2} (G_E^2-1)+\frac{2\tau_p}{(1+\tau_p)^2}(G_MG_E-1)\nonumber\\
&&+\frac{\tau_p^2}{(1+\tau_p)^2}(G_M^2-1)
\end{eqnarray}
into terms of the Sachs form factors. We also note that the eTPE contribution takes into account two structure effects, which are a non-zero value of the anomalous magnetic
moment of the proton $F_2(0)=G_M(0)-1=\kappa\simeq 1.792\,85$ and the finite size of the proton. It is helpful to disentangle them because the former can be found with a very high accuracy, while the latter as we will show (see also \cite{efit}) needs for its calculation additional experimental data and is not free from uncertainties.

Eventually, we arrive at%
%\footnote{\vgi{3rd equation corrected}}
\begin{eqnarray}\label{eq:new:rec}
I_{\rm rec}&=&I_{\rm \kappa}+I_{\rm EF}+I_{\rm M1}+I_{\rm M2}\nonumber \\
I_{\rm \kappa}&=&\kappa\int_0^\infty
{\frac{dq}{q^4}}\bigl\{(2+\kappa)f_{M1}+ f_{M2}\bigr\}\;,\nonumber \\
I_{\rm EF}&=&\int_0^\infty
{\frac{dq}{q^4}}f_{EF}(m,M;q^2)\left[\left(G_E(q^2)\right)^2-1\right]\;,\nonumber\\
I_{\rm M1}&=&\int_0^\infty
{\frac{dq}{q^4}}f_{M1}\left[\left(G_M(q^2)\right)^2-(1+\kappa)^2\right]\;,
\nonumber \\
I_{\rm M2}&=&\int_0^\infty
{\frac{dq}{q^4}}f_{M2}\left[G_M(q^2)G_E(q^2)-(1+\kappa)\right]
\;,
\end{eqnarray}
where
\begin{eqnarray}\label{Eq:tpe:f:F}
f_{\rm EF}&=&f_{E}+\frac{1}{(1+\tau_p)^2}\,f_{F}\;,\nonumber
\\
f_{\rm M1}&=&f_{M}+\frac{\tau_p^2}{(1+\tau_p)^2}\,f_{F}\;,
\nonumber \\
f_{\rm M2}&=&\frac{2\tau_p}{(1+\tau_p)^2}\,f_{F}
\;.
\end{eqnarray}

\begin{figure}[htbp]
\begin{center}
\resizebox{0.9\columnwidth}{!}{\includegraphics{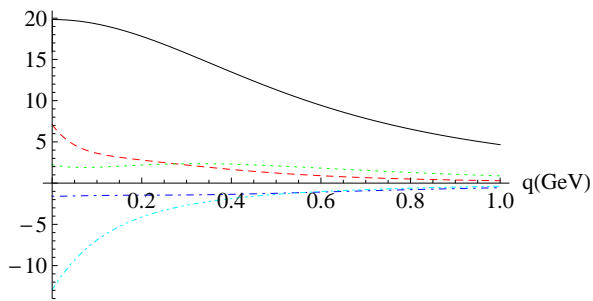}}
\end{center}
\caption{Integrands for the individual contributions to $I_{\rm eTPE}$: for the Friar term (\ref{Eq:friar:1}) (the black line), the $\kappa$ term (the cyan dot-dot-dashed line), the $EF$ term (the red dashed line), the ${M1}$ term (the green dotted line), and the ${M2}$ term (blue dot-dashed line).\label{fig:integrands}}
\end{figure}

We can compare the integrands related to various contributions to $I_{\rm eTPE}$ as summarized in Fig.~\ref{fig:integrands}. The plot
is prepared using the standard dipole approximation
\begin{equation}\label{Eq:G:dip}
G_{\rm dip}(q^2)=\left(\frac{\Lambda^2}{q^2+\Lambda^2}\right)^2
\end{equation}
(with $\Lambda^2=0.71\,{\rm GeV}^2$), which is good to
estimate the size of various contributions, but is not accurate enough
for real calculations.

\section{A preliminary consideration}

As we see from the plot, the Friar term is indeed the dominant contribution. However, that is not only the value of the contribution that matters. The uncertainty is not necessary related to the size of the contribution. The Friar term is not only the dominant term
for the eTPE contribution, but it is also  the part of eTPE, which is the
most sensitive to the parametrization of the form factors. The sensitivity of this term to the parametrization was discussed in \cite{efit}.

\begin{figure}[thbp]
\begin{center}
\resizebox{1.0\columnwidth}{!}{\includegraphics{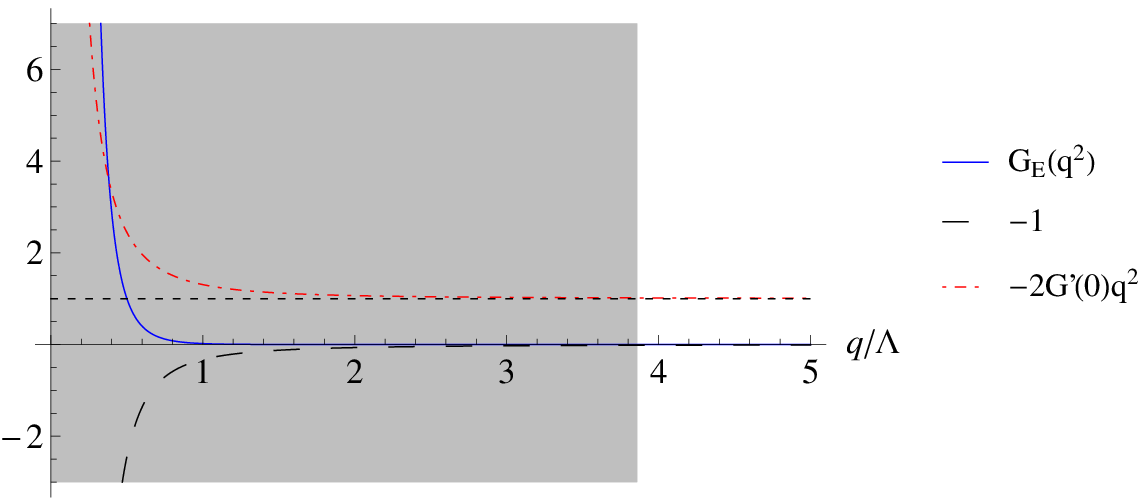}}
\resizebox{0.7\columnwidth}{!}{\includegraphics{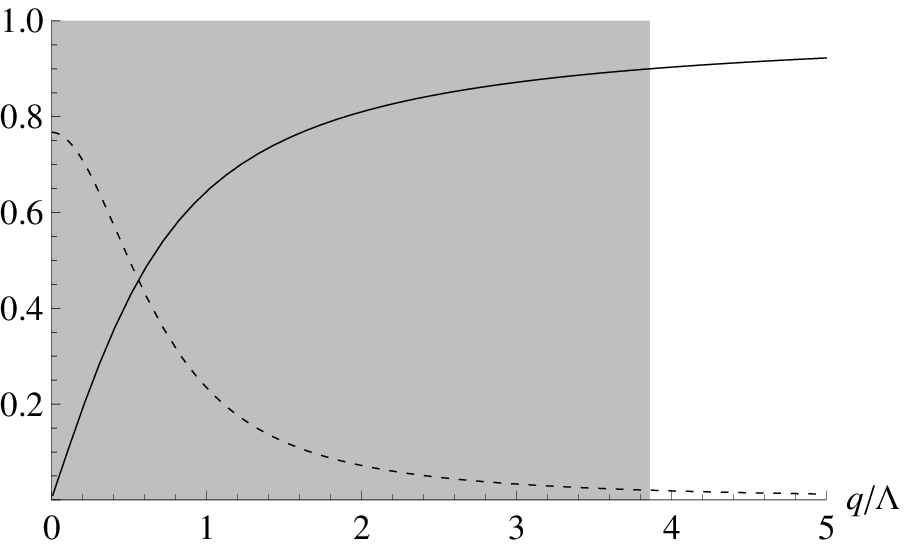}}
\end{center}
\caption{Top: Fractional contributions to the integrand of $I_{\rm Fr}$ in (\ref{Eq:friar:1})
as a function of $q/\Lambda$ as follows from the dipole model. The red
dot-dashed line is for the subtraction term with $G'(0)$. The dashed line is
for the subtraction term with 1 and the blue solid line is for the $G^2$
term, which represents the data. Bottom: the integrand (in arbitrary units) is presented with a dashed line and the integral $\int_0^{q}
{{dq}/{q^4}}\left[(G_E(q^2))^2-1-2G_E^\prime(0)\,q^2\right]$ (in fractional units) in presented as a function of $q$ with a solid line. The shadow belt is for the area in the $q$ space from which 90\% of the integral $I_{\rm Fr}$ come. See \cite{efit} for detail.}
\label{Fig:ind3}       % Give a unique label
\end{figure}

The situation with accuracy of the data for the integration is
presented in Fig.~\ref{Fig:ind3} taken from \cite{efit}. The estimations are made within the dipole parametrization.

The gray area is for the area, where 90\% of the integral comes from. One
sees that below $0.6\Lambda \simeq 0.5\;{\rm GeV}$ there is a
massive cancellation between the data term (with $G_E(q^2)$) and the subtractions terms (with 1 and with $G'(0)$). The data on $G_E(q^2)$ roughly
have accuracy at the 1\% level and below this point the accuracy of the
integrand becomes worse than 1\% and increases very fast. Meanwhile,
above $0.8\Lambda \simeq 0.65\;{\rm GeV}$, the contribution of the
data themselves (i.e. of the $G_E(q^2)$ term) is negligible, while
the subtraction term with $G^\prime_E(0)$ dominates.

Therefore, the integrand at low momentum is determined by our knowledge of the curvature of $(G_E(q^2))^2$, while at high momentum it is determined by the slope of $(G_E(q^2))^2$. Both values cannot be measured experimentally, but can be extracted on base of a certain fitting procedure. Except for the $\kappa$ term,
the integrand of which does not involve the data on the form factors, the other data are also sensitive to the fittings. At low momentum all the integrands are determined by the slope of the related $G^2$ terms and thus are proportional to the electric and/or magnetic radius squared. Indeed, any fitting produces a certain systematic uncertainty.

Unfortunately, the determination of neither electric nor magnetic radius is free of questions. Different extractions from different kinds of data lead to inconsistent results. We have already mentioned a strong discrepancy between muonic-hydrogen $R_E$ \cite{Science} and the scattering values (see, e.g., \cite{mami,sick,zhan,sick2014}). The results on $R_M$ are not in a perfect shape as well (see, e.g., \cite{my_adp,mfit}). However, the discrepancy in determination of $R_M$ from MAMI \cite{mami} and global data \cite{sick} is not that important. The most important is that the determination of the charge radius $R_E$ from the scattering and muonic hydrogen is not in agreement, which compromises the whole procedure. Meanwhile, the magnetic radius is even more sensitive to various systematic effects \cite{comm,resp,sick2012}.

As we mentioned, we discuss a rather small correction. We anticipate that a different treatment of the higher-order proton-size contribution will somewhat shift the value of $R_E$ from muonic hydrogen, but not significantly (cf. \cite{efit}). So, we expect that the discrepancy will remain and we have to find a way how to use the scattering data without a contradiction to the finally extracted value.

We intend to use fits, rather than the data. Generally speaking, to find an appropriate fit is a tricky issue. A
pure empiric fit is the best suited to fit the data, however, it ignores
various theoretical constraints and constraints which follow from some other scattering channels
(through theory). On the contrary, a fit, which follows all those
constraint, either may disagree with the data or may involve so
complicated fit function that it would produce instabilities. Therefore, we
prefer to use empiric fits, but we intend to apply them in a specific way.

There are two ways to apply empiric fits. E.g., when one evaluates the
hadronic vacuum polarization contribution to the anomalous magnetic
moment of muon (see, e.g., \cite{amu:book2}), there is no subtraction involved. The fit is necessary only
as a smooth approximating curve which is consistent with all the data involved.
Its parameters by themselves are irrelevant. The other story is when
the parameters of the fit are really applied. In the integral in
(\ref{Eq:friar:1}) they are. The very shape of an empiric fit has
no physical sense. The parameters must be applied with a caution. A
scenario, in which the result for the Friar correction does not
depend on the low-$q^2$ behavior of the fit, was developed in
\cite{efit}.

The most plausible scenario to explain the discrepancy between the
scattering and muonic-hydrogen values of $R_E$ is that the
scattering data are more or less correct, but the accuracy in
determination of the radius is overestimated. The alternatives, such
as a suggestion that the data are not correct or the muonic value is
wrong, are not an option here. If the muonic value is wrong, there is
no sense in a calculation of small corrections while the discrepancy
is that large. The scattering data from different experiments are
consistent and it is hard to imagine that they are completely wrong.

The integrand involves massive cancellations at low $q$ domain, while for high $q$ it is dominated by the contribution, proportional to
\begin{equation}\label{Eq:G:R26}
G^\prime_E(0)=-\frac16 R_E^2\;.
\end{equation}
If one calculates the integral using scattering data, as it was
done, e.g., in \cite{carlson}, the value of $G^\prime_E(0)$ is indeed
taken from there and it is inconsistent with $R_E$ from the Lamb-shift.
If one calculated the whole integral as a single thing, then there
is no other choice because if we take a value of $G^\prime_E(0)$
inconsistent with the applied fit for $G_E(q^2)$, the integral is
infrared divergent.

Following \cite{efit}, our treatment of the Friar term is based on two important
steps. First, following \cite{efit,pla}, we split the integration into
two parts
\begin{equation}\label{eq:split}
I=\int_0^\infty {{dq}{...}}\equiv I_<+I_>\equiv
\int_0^{q_0}{{dq}{...}} + \int_{q_0}^\infty {{dq}{...}}\;.
\end{equation}
Those two terms are to be treated differently.

The separation disentangles the subtraction at low $q$ and the
asymptotic behavior at high $q$, which both involve
$G^\prime_E(0)$, and, in particular, that allows to use for the
high-$q$ asymptotics the physical value of $R_E$. This value is not fixed, being an adjustable parameter to be eventually found from a
comparison of the theory and experiment on the muonic-hydrogen Lamb
shift. In this area we integrate the $G^2$ term and the subtractions separately and while integrating $G^2$ we rely on the empiric fits.

Meanwhile, at the low-$q$ part of the integral we rely on the expansion of the form factor
\begin{equation}%\label{Eq:low:Fr}
\bigl(G_E(q^2)\bigr)^2\simeq 1 - \frac{R_E^2}{3} q^2 + K\, q^4\;.
\end{equation}
Even a very bold estimation of the $K$ coefficient allows at low $q$ a better accuracy than the data by themselves (see \cite{efit} and consideration below).

The final result takes the form
\begin{equation}\label{Eq:I:hat}
I= C_{R_E^2} \,R_E^2  + \widehat{I}\;.
\end{equation}
Such a form of presentation is crucial, allowing a self-consistent consideration. As the result, the theoretical expression for the Lamb shift in
muonic hydrogen keeps the desirable form as in (\ref{Eq:shape}) and the
higher-order proton-finite-size term (\ref{Eq:friar:1}) contributes
to both coefficients $c_1$ and $c_2$. Meantime, we use the empiric fits only in the area of the high-$q$ part of the integrand and the result only marginally depends on the details of their extrapolation to $q^2=0$  (see \cite{efit} and consideration below). As a final important step, we have a free parameter $q_0$ which we are to vary to minimize the uncertainty of the calculations.

\section{Calculation of the leading proton-finite-size TPE contribution (the Friar term)\label{s:friar:general}\label{s:friar:complete}}

Let's consider evaluation of the leading proton-finite-size TPE contribution $I_{\rm Fr}$ in more details. To deal with a lower-momentum contribution
\[
I_{\rm Fr<}=\int_0^{q_0}
{\frac{dq}{q^4}}\left[\left(G_E(q^2)\right)^2-1-2G_E^\prime(0)\,q^2\right]
\]
we expand the form factor
\begin{equation}\label{Eq:low:Fr}
\bigl(G_E(q^2)\bigr)^2\simeq 1 - \frac{R_E^2}{3} q^2 + K^{\rm dip} (1\pm 1)\, q^4\;,
\end{equation}
and, following \cite{efit}, we estimate the $q^4$ coefficient with help of the standard dipole
model ($K^{\rm dip}=10/\Lambda^4\simeq19.8\;{\rm GeV}^{-4}$). The dipole formula is not that bad for bold estimations. E.g., the
$G'(0)$ terms are different from the value following from the muonic $R_E$
by roughly 7\% and from those following from hydrogen-and-deuterium
spectroscopy's $R_E$ and from scattering's $R_E$ by about 17\%. The
uncertainty of 100\% for the curvature seems reasonable. The value
is also consistent with the curvature of various empiric fits (see,
e.g., \cite{kelly,as2007,am2007,ab2009,va2011,thesis};
see also a discussion below).

The physical condition on the expansion is $q_0<2m_\pi\simeq
0.28\;{\rm GeV}$. We are to vary $q_0$ to minimize the final uncertainty and we will check this constraint after we specify the optimal
value of $q_0$.

The expansion (\ref{Eq:low:Fr}) allows to easily find the low-momentum contribution and its uncertainty
\begin{equation}\label{Eq:low:Fr1}
I_{\rm Fr>}=  10\,(1\pm1) \, \frac{q_0}{ \Lambda^4}\;.
\end{equation}

For the low-$q$ part we have exactly followed \cite{efit}, while for the high-momentum part we prefer a certain alternation, which would be more suitable for the calculation of the recoil corrections. The higher-momentum part contains the data-related and radius-related contributions
\begin{equation}\label{Eq:high:Fr}
I_{\rm Fr>}=\int_{q_0}^\infty
{\frac{dq}{q^4}}\left(G_E(q^2)\right)^2-\frac{1}{3q_0^3}+\frac13\frac{R_E^2}{q_0}\;.
\end{equation}

Combining the low-$q^2$ and high-$q^2$ contributions, we find
%\footnote{\vgi{We need explicit definitions of $C_{R_E^2:\dots}$ somewhere.}}
\begin{eqnarray}\label{eq:decomp:fr}
\widehat{I}_{\rm Fr}&=&
\int_0^{q_0}{\frac{dq}{q^4}}\left[\left(G_E(q^2)\right)^2-1-2G_E^\prime(0)\,q^2\right]
\nonumber\\
&&
+ \int_{q_0}^\infty {\frac{dq}{q^4}}\left[\left(G_E(q^2)\right)^2-1\right]\;,\nonumber\\
C_{R_E^2:{\rm Fr}}&=&\frac1{3q_0}\;,
\end{eqnarray}
which is necessary for the decomposition according to (\ref{Eq:I:hat}).

To calculate the central value of the data-related term in
(\ref{Eq:high:Fr}), the integration in \cite{efit} was done over the
fits \cite{kelly,as2007,am2007,ab2009,va2011} (see Appendix~\ref{s:efit}) and as the result the
related median value was taken. Here, we use the most recent empiric fit from \cite{va2011}
\onecolumngrid
%~\\~\\
\begin{equation}\label{fit:va2011}
G_E(q^2)=
\frac{1+2.909\,66\tau_p-1.115\,422\,29\tau_p^2+3.866\,171\times10^{-2}\tau_p^3}{1+14.518\,7212\tau_p+40.883\,33\tau_p^2+99.999\,998\tau_p^3 +4.579\times10^{-5}\tau_p^4+10.358\,0447\tau_p^5}\;,
\end{equation}
%~\\
\twocolumngrid
\noindent
where $\tau_p$ is defined above in (\ref{Eq:tau:p}). We estimate our knowledge of the form factors as at the level of 1\% and expect that the dominant contribution to the integral for $G^2$ comes from a relatively small area around the low limit (cf. \cite{efit}). So, we expect the whole $G^2$ contributions known at the level of 2\% and thus we arrive at
\onecolumngrid
%~\\~\\
\begin{equation}\label{Eq:high:Fr2}
I_{\rm Fr>}=\int_{q_0}^\infty
{\frac{dq}{q^4}}\left(G_E(q^2)\right)^2\times\bigl(1\pm0.02\bigr)-\frac{1}{3q_0^3}+\frac13\frac{R_E^2}{q_0}\;,
\end{equation}
%~\\
\twocolumngrid
\noindent
where $G_E(q^2)$ is taken from (\ref{fit:va2011}).

While choosing the fits, we have looked
for those which produce a smooth function for the all space-like
$q^2$ and have a good value of the reduced $\chi^2$. We prefer
`simple' empiric fits. We believe that it is tricky to determine the
accuracy for the extrapolation to low $q^2$ and for the subsequent
differentiation (to find $R_E$) and,  most probably, their
inconsistency with the muonic value of the proton radius is due to an
overestimation of their accuracy. Nevertheless, the form factor by
itself (not its derivative) is fitted by those fits well. The
accuracy of the application of the fits was estimated in \cite{efit}
separately as
\begin{equation}\label{Eq:high:Fr1}
\delta I_{\rm Fr>}=  \frac{1}{3q_0^3}\frac{2\delta G_E(q_0^2)}{G_E(q_0^2)}\,\bigl(G_{\rm dip}(q_0^2)\bigr)^2\;,
\end{equation}
where we expect that $\delta G_E(q_0^2)/G_E(q_0^2)\simeq 1\% $.

Here, we apply a somewhat different estimation of the uncertainty. It is estimated as 2\% of the $G^2$ contribution into $I_{\rm Fr>}$. The dominant part of the $G^2$ integral comes from the low end of the integration area and as far as the low-end area dominates the estimations are close.

Above we explained that it is not consistent to use any fits. However,
we need to distinguish using of the central value of the fits with a
pre-estimated uncertainty, as we do, and the use of the parameters of the
fits to find, e.g., some derivatives etc. If the fit over a large
sample of the data has the reduced $\chi^2$ about unity, that means
that the fit is consistent with the data at the level of their
uncertainty, which is about 1\%. Such use of the fit does not assume
any model. Once one intends to take advantages of the large number of
data points and to obtain a certain value, e.g., $G'(0)$ with
a high precision because of statistical average, the model comes out
and the systematic uncertainty due to its application is to be
estimated. We prefer to stay on the safe side and introduce the
uncertainty at the level of the accuracy of the data which grants
model independence while applying any fit with a good $\chi^2$
value.

The scatter due to choice of the parametrization (i.e. due to choice of
the fits \cite{kelly,as2007,am2007,ab2009,va2011}) is smaller than
the uncertainty we introduced. The reason for the bigger value
of the uncertainty is
that all the fits basically fitted the same data, which were only partly
updated from a publication to a publication. The scatter does not reflect the accuracy of the data, but rather
the importance of their updates. The uncertainties of $I_{\rm Fr<}$ and
$I_{\rm Fr>}$ are considered as independent.

\begin{table}[htbp]
\begin{center}
\begin{tabular}{l|c|c|c|c}
\hline
~Fit & Ref. & Type &  $R_E$ [fm] &  ~$K$ [GeV$^{-4}$]\\[0.8ex]
\hline
(\ref{fit:as2007}) & \cite{as2007}& chain fraction & 0.90 & 34.3 \\[0.8ex]
(\ref{fit:bm2005}) & \cite{as2007}& chain fraction & 0.90 & 35.3 \\[0.8ex]
(\ref{fit:kelly}) & \cite{kelly}& Pad\'e approximation ($q^2$) & 0.86 & 28.0 \\[0.8ex]
(\ref{fit:am2007}) & \cite{am2007}& Pad\'e approximation ($q^2$) & 0.88 & 31.1 \\[0.8ex]
(\ref{fit:ab2009}) & \cite{ab2009}& Pad\'e approximation ($q^2$) & 0.87 & 28.2 \\[0.8ex]
(\ref{fit:va2011}) & \cite{va2011}& Pad\'e approximation ($q^2$) & 0.88 & 31.3 \\[0.8ex]
\hline
\end{tabular}
\caption{The low-momentum expansion of the fits for the electric
form factor of the proton applied in \cite{efit}. The values are
given for central values of the fits without any uncertainty. Here:
$\bigl(G_E(q^2)\bigr)^2= 1 - R_E^2 q^2/3 + K q^4 + ...$. The related
values for the standard dipole fit are $R_E=0.811\;$fm and
$K=19.8\;{\rm GeV}^{-4}$. The spread of the central values of the
charge radius (from 0.86 to 0.90\;fm) for different fits (from different sets of the data) is comparable with the spread of
central values of the updated results on the proton radius, obtained by different methods, the central values
of which are 0.84\;fm ($\mu$H \cite{Science}), 0.88\;fm (H{\&}D
\cite{codata2010}), 0.895\;fm (global data without MAMI
\cite{sick}---and the results of the other evaluations are similar)
and 0.88\;fm (MAMI \cite{mami}). All the values, but from muonic
hydrogen, have uncertainty at the level of 1\%. The muonic one is
much more accurate. \label{t:exp:fits}}
 \end{center}
 \end{table}

To facilitate further numerical evaluations with including the
recoil effects, we choose here instead of the median  the fit
\cite{va2011}, which is the most recent one. This modification is
not important because our estimation of the uncertainty is
substantially larger than the scatter. The most of the integral (for
the $G^2$ term) comes from the low limit. Here, we choose to assign
the 2\% uncertainty to the whole $G^2$ term. We have checked that
this modification is of marginal importance for the Friar term,
however, as we mentioned, it simplifies a comparison with \cite{carlson}.
Note, that is not an improvement, but a simplification. The reason
to check the median using various fits (some of which are partly out of
date) was to check that the result does not much depend on low-$q^2$
behavior. The spread of the values of the radius was sufficient to prove
such  model independence, which was also confirmed by the use a fit from
\cite{bo1994}, which, despite of unreasonable low-$q$ behavior,
produces a quite reasonable result. Only  after such tests are performed,
the use of
the most recent fit is possible.

\begin{figure}[thbp]
\begin{center}
\resizebox{1.0\columnwidth}{!}{\includegraphics{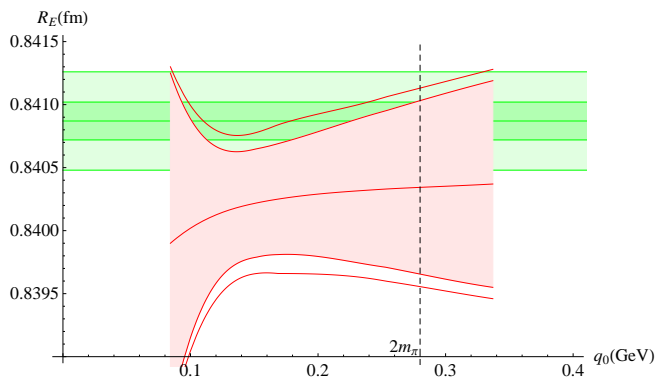}
}
\end{center}
\caption{The value of the proton charge radius $R_E$ extracted
\cite{efit} from measurement \cite{Science} of the muonic-hydrogen
Lamb shift. The plot follows \cite{efit}. The horizontal belt is for the original value of the
parameter. The other figure [of a more complicated shape] is for the
radius determined in \cite{efit}, presented as a function of the
separation parameter $q_0$. Every area corresponds to one sigma. The
internal and darker part of each figure is related to the partial
uncertainty due to the calculation $I_{\rm Fr}$, while the whole
area is for the total uncertainty of the radius. The final result of
\cite{efit} $R_E=0.840\,22(56)\;$fm was obtained there at the
optimal value $q_0\simeq0.152\;{\rm GeV}$.}\label{Fig:rscience-q0}       % Give a unique label
\end{figure}

The extracted value of the proton radius depends on the separation parameter
$q_0$ in almost the same way as for \cite{efit}. The procedure (see \cite{efit} and sections below) allows to vary
the value of the separation parameter $q_0$ and to find an optimal
one. The results with different values of $q_0$ are consistent. The deviation
from the former treatment of the eTPE \cite{carlson} is large enough (at the
level of the uncertainty) to be considered seriously. The dependence $R_E(q_0)$ is shown in Fig. \ref{Fig:rscience-q0}.

Minimizing the uncertainty we arrive at
\begin{equation}\label{Eq:result2}
I_{\rm Fr}= \frac13\frac{R_E^2}{q_0}-23.5(3.3)\;{\rm GeV}^{-3}\;,
\end{equation}
where the numerical value corresponds to $q_0\simeq0.147\;{\rm GeV}$. It is important that the new optimal value is quite below $2m_\pi$ which is the
requirement for the validity of the expansion (\ref{Eq:low:Fr}).

While the decomposition $R^2$ term and into the constant $\widehat{I}$ is somewhat different\footnote{The result of \cite{efit} is $I_{\rm Fr}= {R_E^2}/{(3q_0)}-22.2(3.4)\;{\rm GeV}^{-3}$, where the numerical value uses the optimal value $q_0\simeq0.152\;{\rm GeV}$. (Indeed, since the uncertainty of $I_>$ for the Friar term is defined here and in \cite{efit} differently, the optimal values, which minimize the uncertainty, are also somewhat different.)}, the related numerical results at $R_E=0.84\;{\rm fm}$ are in a perfect
agreement, being $17.6(3.4)\;{\rm GeV}^{-3}$ \cite{efit} and
$17.7(3.3)\;{\rm GeV}^{-3}$ (this work). The approach applied here is a modification of the one from \cite{efit} and the results for the radius are expected to nearly coincide at $R_E=0.84\;{\rm fm}$. The latter effectively means that the extracted radius is to be approximately the same (with a marginal deviation). (We widely use in this paper the `reference' value of various contributions at $R_E=0.84\;{\rm fm}$---that is the approximate value consistent with extractions from the muonic hydrogen \cite{Science,efit} at the level of $10^{-3}$ which is completely sufficient taking into account the uncertainty of our calculations of the Friar and eTPE integrals at the level of 20\%.)

Those numerical values should be
compared with the results which can be obtained if one directly applies empiric fits to (\ref{Eq:friar:1}). The related results are
%$17.2\;{GeV}^{-3}$ (from the stanesdard dipole fit \ref{Eq:G:dip}),
$I_{\rm Fr}=23.1(1.1)\;{\rm GeV}^{-3}$  (after applying empiric fits
\cite{Fri:sick}), $22.9(1.2)\;{\rm GeV}^{-3}$ (after applying
\cite{efit} empiric fits \cite{am2007} and \cite{kelly} --- that is
similarly to consideration in \cite{carlson}, which was done for the
complete $I_{\rm eTPE}$) and $I_{\rm Fr}=24.3(7)\;{\rm GeV}^{-3}$ (after
applying empiric fits \cite{Fri:mami}). The uncertainty of our result is essentially larger than theirs and the the shift from the quoted values to ours is larger than the uncertainty. We believe that the uncertainty of the cited results was underestimated and suffered from an inconsistency (once we accept the muonic value for the proton radius).

We have to recall the geometrical meaning of the Friar term
\begin{equation}\label{Eq:friar:coord}
I_{\rm Fr}=\frac{\pi}{48}\int{d^3r}\int{d^3r'}|{\bf r}-{\bf r}'|^3\rho({\bf r})\rho({\bf r}')\;,
\end{equation}
where in a non-relativistic approximation $\rho({\bf r})$, being a Fourier transform of $G_E({\bf q})$, is the charge density inside the proton.
The muonic-hydrogen data assume a smaller
proton, thus the Friar term adjusted to those data should also be
smaller (see also discussion on that in \cite{Fri:mami}). The
radius is smaller by about 6\% than the results from the empiric
fits and the Friar term (the central value) is smaller by about
15\%, which seems consistent.

Indeed, in a rigorous consideration, the Fourier transform of $G_E$ is not the charge density in common sense. However, the presence of relativistic effects does not affect the fact that the value of the Friar term is determined to a certain extent by the proton size, it just makes this relation more complicated and involves corrections. The non-relativistic part with a clear geometric sense remains and it is dominant.

\begin{figure}[thbp]
\begin{center}
\resizebox{0.8\columnwidth}{!}{\includegraphics{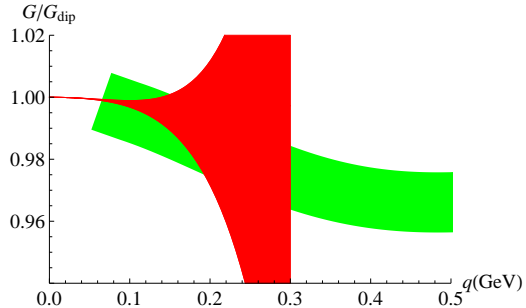} }
\end{center}
\caption{The continuity test for the proton electric form factor normalized against the standard dipole fit. The red area on the left with fast increasing (to the right) width is for the low momentum approximation, while the belt with approximately homogenous width on the right is for the empiric fit \cite{va2011} we used, with the uncertainty of 1\%.}
\label{fig:cont}       % Give a unique label
\end{figure}

One may wonder whether
splitting the area of the integration maintains the
continuity of the form factor. We cannot check it {\em a priori\/}
because at low $q^2$ we use the actual value of the proton radius.
However, once the complete calculation is performed, we find the
value of the radius and can to check the continuity. Such an {\em a
posteriori\/} test is plotted in Fig.~\ref{fig:cont}. The red area
is for the low-$q^2$ expansion and the green one is the fit from
\cite{va2011} suggesting the uncertainty of 1\%. They are
consistent.
%Here, we expect a marginal shift of the value of the radius and use for the plot the value from \cite{efit}.

\begin{figure}[htbp]
\begin{center}
\resizebox{0.9\columnwidth}{!}{\includegraphics{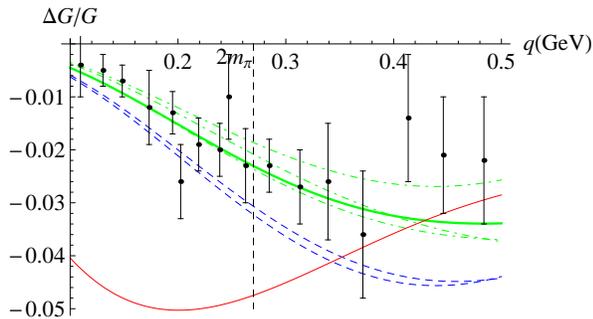}}
\end{center}
\caption{The fractional deviation of the fits (see in App.~\ref{app:efit} for details) and the data for the
electric form factor $G_E(q^2)$ from the standard dipole form
factor. The data are taken from \cite{am2007}, where the proton form
factors were extracted from experimental data on the elastic
electron-proton scattering. The fits include the chain-fraction fits (\ref{fit:as2007}) and (\ref{fit:bm2005})
(in $q^2$) \cite{am2007} (the blue dashed line), the
Pad\'e-approximation (in $q^2$) fits (\ref{fit:kelly}), (\ref{fit:am2007}), (\ref{fit:ab2009}), and (\ref{fit:va2011}) \cite{kelly,am2007,ab2009}
(the green dot-dashed lines) with the most recent one from \cite{va2011}
shown with a green bold solid line. An exceptional fit (\ref{fit:bo1994}) with an inverse polynomials in $q$ \cite{bo1994} is shown with a solid red
line. \label{fig:nnn}}
\end{figure}

The method suggested in \cite{efit} pretends that the result does
not depend on the parametrization of the fit. To check that, a number
of fits have been applied (see Fig.~\ref{fig:nnn}).  Those include
the chain-fraction fits
(in $q^2$) \cite{am2007} (the blue dashed line), the
Pad\'e-approximation (in $q^2$) fits \cite{kelly,am2007,ab2009}
(the green dot-dashed lines) and \cite{va2011} (the green bold solid
line), and a fit with the inverse polynomials in $q$ \cite{bo1994} (the
solid red line). The fits except for \cite{bo1994} have reasonable
low $q$ behavior. They are analytic in $q^2$ around $q^2=0$ and the
related value of the charge radius varies from 0.86 to 0.90~fm, so
the spread is comparable with the distance to the muonic value of
0.84 fm (see Table~\ref{t:exp:fits}). The high-$q^2$ asymptotics is
proportional to $q^{-2}$ for
the chain fractions and to $q^{-4}$ for the Pad\'e approximations.
Therefore, the parameters and the general behavior (far from the
characteristic $q_0$ area) vary very much. Nevertheless, the spread
of the values for the integral $I_{\rm Fr>}$ is smaller than the estimation
of the uncertainty \cite{efit}. The scatter of the results with
inclusion of the fit with unrealistic behavior \cite{bo1994} is somewhat larger
than the uncertainty but still acceptable \cite{efit}.

\section{Calculation of the recoil corrections}

After discussing the leading eTPE term (within the external field approximation), we are prepared to consider the eTPE recoil corrections to it. We start from (\ref{eq:new:rec}). As we see from Fig.~\ref{fig:integrands}, the dominant proton-finite-size terms are the $EF$ and the $\kappa$ ones. The former is the proton-size term, while the latter is not related to the proton size directly.
Domination of the $EF$ term is due to $q/M$ expansion. Comparing the coefficient functions at the area of $m\ll q\ll M$ we find
\begin{eqnarray}\label{Eq:tpe:f:lead:new}
f_{\rm EF}(m,M;q^2)&\simeq&-\frac{3}{8}\frac{q}{M}\;,\nonumber \\
f_{\rm M1}(m,M;q^2)&\simeq&-\frac{3}{16} \left(\frac{q}{M}\right)^3   \;,
\nonumber \\
f_{\rm M2}(m,M;q^2)&\simeq&\frac{3}{16}\left(\frac{q}{M}\right)^3   \;,
\end{eqnarray}
which explains why the $EF$ term is dominating.

An important area of the integration is related to $q\sim m$ and it is useful to present a more detailed asymptotics at $m \ll M$, $q\ll M$ (exactly in $m/q$)
\onecolumngrid
%~\\~\\
\begin{equation}
f_{\rm EF} =-\frac{q^4\left(\sqrt{4m^2+q^2}-q\right)+ 8m^4\left(\sqrt{4m^2+q^2}-2m\right)- 2m^2q^3}{16m^4M}+\dots
\end{equation}
%~\\
\twocolumngrid

%\subsection{EF}

Decomposing the $EF$ term according (\ref{Eq:I:hat}) we find
\begin{eqnarray}\label{eq:decomp:ef}
\widehat{I}_{\rm EF}&=&
\int_0^{q_0}{\frac{dq}{q^4}}\left[\left(G_E(q^2)\right)^2-1-2G_E^\prime(0)\,q^2\right]f_{\rm EF}(q^2)
\nonumber\\
&&
+ \int_{q_0}^\infty {\frac{dq}{q^4}}\left[\left(G_E(q^2)\right)^2-1\right]f_{\rm EF}(q^2)\;,\nonumber\\
C_{R_E^2:{\rm EF}}&=&-\frac13\int_0^{q_0}{\frac{dq}{q^2}}f_{\rm EF}(q^2)\;.
\end{eqnarray}
That is very similar to the expression (\ref{eq:decomp:fr}) for the Friar term. The evaluation of the integrals is done in a similar way and we apply the same effective description for $G_E$. The only difference is that in former case
all the integrals except for the $G^2$ term were calculated analytically,
while here we have to do that numerically.

The other important term is the $\kappa$ term. It is a proton-structure correction, but not a proton-finite-size one. Since we decompose elastic TPE into the point-like Salpeter term and a proton-structure term, the subtraction in the latter is due to a proton without any anomalous magnetic moment. Therefore, there is a contribution to eTPE which does not vanish in the limit $R_E=R_M=0$, which is the $\kappa$ term. It does not involve the parametrization of the form factors and may be found with a high accuracy. We obtained the result from both numerically
\begin{equation}
I_{\kappa}=-2.850\,23\;{\rm GeV}^{-3}
\end{equation}
and in a close analytic form
\begin{eqnarray}%\label{prosto2}
I_{\kappa}&=& \frac{\kappa}{16\,M^2(M-m)}\cdot
\left[-3(1+\kappa)\ln{\frac{M}{m}}\right.\nonumber\\
&&\left.+(1-\kappa)\left(\ln{2}-\frac14\right)+B\cdot\left(\frac{m}{M}\right)^2\right]\;.
\end{eqnarray}
where the value of $B\sim{\cal O}(1)$. Our result for this value is
\begin{eqnarray}
B&=&\left[\frac{\kappa-2}{4}\ln{\frac{2M}{m}}+\frac{2+11\kappa}{48}\right]\nonumber\\
&&+\left[\frac{\kappa-3}{16}\ln{\frac{2M}{m}}-\frac{43(1+\kappa)}{384}\right]\left(\frac{m}{M}\right)^2
+\dots\nonumber
%{\cal O}\left(\left(\frac{m}{M}\right)^4\right)\;.
\end{eqnarray}
As one sees (cf. (\ref{Eq:tpe:f:lead:new})), this term is $\sim M^{-3}$ in contrast to the leading part of the $EF$ term, which is $\sim M^{-1}$. However, the non-vanishing value of $\kappa$, which enters the combinations such as $\kappa(1+\kappa)$, produces a certain enhancement.

Let's now consider the magnetic finite-size terms, $I_{\rm M1}$ and $I_{\rm M2}$. At low-$q$ we use expansion
\begin{equation}
\bigl(G_M(q^2)\bigr)^2=\mu_p^2\left(1-\frac{R_{\rm dip}^2}{3}q^2[1\pm0.4]+ K^{\rm dip}q^4(1\pm1)\right)\,,
\end{equation}
where $\mu_p=1+\kappa_p\simeq 2.792\,85$ is the proton magnetic moment
in the units of the nuclear magnetons, for the magnetic form factor, similar to that for the electric one. However, there is a difference. While for the
electric form factor the value of $R_E$ is the adjustable parameter for the extraction of the charge radius from muonic hydrogen, the rms magnetic radius $R_M$ should be a constraint. We use here the dipole value of the radius
\[
R_{\rm dip}^2=\frac{12}{\Lambda^2}\;,
\]
but with a conservative uncertainty of 40\% (to $R_M^2$).

As for the high-$q$, we apply the fits from \cite{va2011}, which gives not only the fit for the electric form factor (\ref{fit:va2011}), but also a fit for the magnetic one. The latter
is of the form
\onecolumngrid
%~\\~\\
\begin{equation}\label{fit:mva2011}
\frac{G_M(q^2)}{\mu_p}=\frac{1-1.435\,73\tau_p +1.190\,520\,66\tau_p ^2+0.254\,558\,41\tau_p ^3}{1+9.707\,036\,81\tau_p +3.7357\times10^{-4}\tau_p ^2+6.0\times10^{-8}\tau_p ^3+ 9.952\,7277\tau_p ^4+12.797\,7739\tau_p ^5}\;.
\end{equation}
%~\\
\twocolumngrid

The decomposition into $\widehat{I}$ and $C_{R_E^2}$ and further numerical
evaluations of those values are similar to those for the $EF$ term.

The numerical results for eTPE as a function of different values of the
separation parameter  are summarized in Table~\ref{t:ivenkat}. The minimization of the
uncertainty leads to $q_0\simeq0.147\;{\rm GeV}$ as the optimal value.

\begin{table}
\begin{tabular}{l|c|c|c}
\hline
$q_0\;[{\rm GeV}]$ & $C_{R_E^2}\;[{\rm GeV}^{-1}]$ & $\widehat{I}_{\rm eTPE}\;[{\rm GeV}^{-3}]$ & $I_{\rm eTPE}\;[{\rm GeV}^{-3}]$\\
\hline
$0.1$ & $3.362\,20$ &  $-44.52$ & $16.4(6.0)$\\[0.8ex]
$0.2$ & $1.714\,65$ & $-12.79$ & $18.3(4.0)$\\[0.8ex]
$0.3$ & $1.175\,25$ & $-2.597$ & $18.7(5.8)$\\[0.8ex]
$0.4$ & $0.911\,53$ & $2.527$ & $19.1(7.6)$\\[0.8ex]
$0.5$ & $0.757\,09$ & $5.790$ & $19.5(9.2)$\\[0.8ex]
\hline
\end{tabular}
\caption{The results $\widehat{I}$ and $I_{\rm eTPE}$ for fits (\ref{fit:va2011}) and
(\ref{fit:mva2011}) at various $q_0$. The last column is for the numerical value of  $I_{\rm eTPE}$ at $R_E=0.84\;$fm.
\label{t:ivenkat}
}
\end{table}

The result is
\begin{equation}
I_{\rm eTPE}=\bigl[-24.14(3.27) + 59.31\,r_p^2\bigr]\;{\rm GeV}^{-3}\;,
\end{equation}
where the error of $2.90\;{\rm GeV}^{-3}$ corresponds to $\widehat I_{\rm eTPE<}$, while the error of $1.53\;{\rm GeV}^{-3}$ relates to $\widehat I_{\rm
eTPE>}$.

\begin{figure}[htbp]
\begin{center}
\resizebox{0.9\columnwidth}{!}{\includegraphics{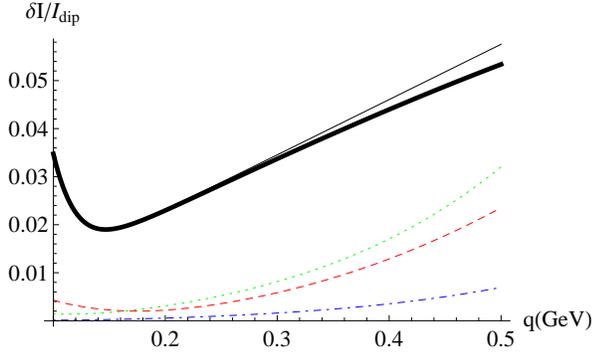}}
\end{center}
\caption{
Individual contributions to the uncertainty $\delta I_{\rm eTPE}$ in relative units of the dipole value for various $q_0$.
The total error $\delta I$ is presented by the thick solid line,
other types of lines are described in the caption of Fig.~\ref{fig:integrands}. The recoil contributions are very small and to put all the contributions into the same picture we use different normalization. The values of $\delta I_{\rm eTPE}$ and $\delta I_{\rm Fr}$ are multiplied by 0.1, which allows to use a linear scale (instead of the logarithmic one). The recoil corrections are [anti] correlated with the Friar term, which somewhat reduces the total uncertainty.
\label{fig:f2}
}
\end{figure}

The individual contributions to the eTPE correction are presented in Table~\ref{t:indiv}, while the individual contributions to the uncertainty are summarized in Fig.~\ref{fig:f2}.
We see that the individual electric and magnetic recoil
contributions are at the level of 10-15\%, however, they have a trend
for a strong cancellation. The cancellation is rather accidental and depends
on the value of the anomalous magnetic moment of the proton
$\kappa_p$, the mass ratio $m/M$ and the proton radius. The terms differently depend on the parameters and the
cancellation of the central values does not produce a massive
cancellation of the uncertainties.

However, the latter are much smaller, than the level of the 10\% of the Friar-term uncertainty. The size of the contribution is not that important as its sensitivity to the parametrization. It is also important, that any uncertainties at the level of 10\% of the Friar-term uncertainty would be important only if it is correlated to the main uncertainty and we have to sum them up linearly rather than as the rms sum. That leaves us with the $EF$ term as the main source of such
an uncertainty since it is expressed in terms of the same parameters as the leading term. The contribution to the integral by itself can come from a various area of $q$, however, the area sensitive to the parametrization is around $q_0$. The expansion at low $q$ ($<q_0$) is the most uncertain at the largest `low $q$' available, while the uncertainty of the high-$Q$ area ($>q_0$) is mostly determined by the smallest `high $q$' available. The contribution comes from a more broad area than that around $q\sim q_0\simeq 0.147\;{\rm GeV}$. In the limit $m\ll q_0\ll M$, the suppression of $EF$ behaves as $q_0/M$ with small coefficients ($<1$) as follows from the asymptotic behavior (\ref{Eq:tpe:f:lead:new}). The optimal $q_0$ is found to be $q_0\simeq 0.147\;{\rm GeV}$ and thus $q_0<2m_\mu$. That leads to a suppression of the $EF$ term in respect to the $Fr$ term by a factor of $m/M$ ($\sim 10\%$) with small coefficients ($<1$) partly because of additional softening of the integrands by factors $q/2m_\mu<1$ and softening of the character of the $q$-integration.

\begin{table}
\begin{tabular}{l|c|c|c}
\hline
Contribution & $C_{R_E^2}\;[{\rm GeV}^{-1}]$ & $\widehat{I}_{\rm eTPE}\;[{\rm GeV}^{-3}]$ & $I_{\rm eTPE}\;[{\rm GeV}^{-3}]$\\
\hline
Fr & $2.271$ & $-23.5(3.3)$ & $17.68(3.31)$ \\[0.8ex]
$\kappa$  & $0$ & $-2.850$ & $-2.850$ \\[0.8ex]
EF & $0.039$ & $1.26(4)$ & $1.97(4)$ \\[0.8ex]
M1 & $0$ & $1.76(3)$ & $1.76(3)$ \\[0.8ex]
M2 & $-0.0006$ & $-0.838(4)$ & $-0.849(4)$ \\[0.8ex]
\hline
eTPE & $2.309$ & $-24.1(3.3)$ & $17.71(3.27)$ \\[0.8ex]
\hline
\end{tabular}
\caption{Individual contributions to the $I_{\rm eTPE}$ at optimal $q_0\simeq0.147\;{\rm GeV}$. The last column is for numerical value of $I_{\rm eTPE}$ at $R_E=0.84\;$fm.
\label{t:indiv} }
\end{table}

As an important test we check that the scatter due to use of the empiric fits (except for those from \cite{bo1994}) is below the uncertainty. The fits from \cite{bo1994} shift the result somewhat farther, but the value is acceptable because of the unreasonable behavior of the fits at low $q$.

\begin{table}
\begin{tabular}{l|l|l|c|c}
\hline
Fit & Ref. & Type & $\widehat{I}$ & $I_{\rm eTPE}$\\
 &  &  & $[{\rm GeV}^{-3}]$ & $[{\rm GeV}^{-3}]$\\
\hline
%(A0)                                   &               & dipole  & $-21.85$ & $20.00$ \\[0.8ex]  %Dip
(\ref{fit:as2007}), (\ref{fit:mas2007}) & \cite{as2007}& chain fraction               & $-24.86$ & $16.99$ \\[0.8ex]  %Arr1
(\ref{fit:bm2005}), (\ref{fit:mbm2005}) & \cite{as2007}& chain fraction               & $-25.01$ & $16.83$ \\[0.8ex]  %Arr2
(\ref{fit:kelly}), (\ref{fit:mkelly})   & \cite{kelly} & Pad\'e approx. ($q^2$) & $-23.66$ & $18.18$ \\[0.8ex]  %Kelly
(\ref{fit:am2007}), (\ref{fit:mam2007}) & \cite{am2007}& Pad\'e approx. ($q^2$) & $-24.18$ & $17.66$ \\[0.8ex]  %Arr3
(\ref{fit:ab2009}), (\ref{fit:mab20091})& \cite{ab2009}& Pad\'e approx. ($q^2$) & $-23.86$ & $17.98$ \\[0.8ex]  %Alb
(\ref{fit:ab2009}), (\ref{fit:mab20092})& \cite{ab2009}& Pad\'e approx. ($q^2$) & $-23.87$ & $17.98$ \\[0.8ex]  %Alb
(\ref{fit:va2011}), (\ref{fit:mva2011}) & \cite{va2011}& Pad\'e approx. ($q^2$) & $-24.14$ & $17.71$ \\[0.8ex]  %Venkat
(\ref{fit:bo1994}), (\ref{fit:mbo1994}) & \cite{bo1994}& inv. polynom. ($q$)   & $-29.31$ & $12.54$ \\[0.8ex]  %Bosted
\hline
\end{tabular}
\caption{Numerical results for central values of $\widehat{I}$ and $I_{\rm eTPE}$ at the ``optimal'' $q_0\simeq0.147\;{\rm GeV}$ from various fits. The last column is for numerical value of  $I_{\rm eTPE}$ at $R_E=0.84\;$fm.
% For the dipole fit $\Lambda^2=0.71{\rm GeV}^2$ is assumed.
\label{t:ifits}
}
\end{table}

%\newpage

\section{Concluding remarks}

The extraction equation is of the similar form as for the Friar term in \cite{efit}
\onecolumngrid
%~\\~\\
\begin{equation}\label{eq:extract4}
(R_E)^2-(R_E^{(0)})^2 =\frac{24(Z\alpha)m_r}{\pi}\frac{1}{1-\frac{24(Z\alpha)m_r}{\pi}C_{R_E^2}(q_0)}\biggl[C_{R_E^2}(q_0)(R_E^{(0)})^2
+ \,\left(\widehat{I}_{\rm eTPE}(q_0)-I_{\rm eTPE}^{0}(q_0)\right)\biggr]\;,
\end{equation}
%~\\
\twocolumngrid
\noindent
where $I_{\rm eTPE}^{0}$ is the original result for $I_{\rm eTPE}$
and $\widehat{I}_{\rm eTPE}$ is the one from which we removed the $R_E^2$ term. The
coefficient $C_{R_E^2}$ is due to a small `renormalization' of the $R^2_E$ term.

In the case of the Friar term the `renormalization' is
\[
1\to 1-\frac{8(Z\alpha)}{\pi}\frac{m_r}{q_0}
\]
and the recoil effects weakly affect it. For $I_{\rm eTPE}^{0}$ we
use here the result from \cite{carlson} with the $F_1^2$ term restored as suggested in \cite{birse}
% \footnote{Note, actually the value has not been calculated rigorously. Originally, a certain method to find the central value of the eTPE correction and its
% uncertainty was suggested in \cite{carlson}. The $F$ term was ignored there. However, its central value was given in the discussion section of \cite{carlson}
% and latter on it was added up in \cite{birse} and subsequent citations. Its uncertainty was never discussed. Since the uncertainties are correlated they sum up % rather linearly and as the rms sum. Even the small uncertainties should be taken into account.}%\footnote{\vgi{$I_{\rm eTPE;(A5)+(A6)}^{0}=(25.9-4.4)\;{\rm
% GeV}^{-3}=21.6\;{\rm GeV}^{-3}$}}
\begin{eqnarray}
I_{\rm eTPE}^{0}&=&\Bigl(27.5(1.6)-4.5\Bigr)\;{\rm GeV}^{-3}\nonumber\\
&=&23.0(1.6)\;{\rm GeV}^{-3}
\;.\end{eqnarray}
That is related to the evaluation in \cite{Science}.

Eventually, we arrive at
\begin{equation}\label{eq:re:result}
R_E=0.840\,22(53)\;{\rm fm}\;,
\end{equation}
which is to be compared with the original result $R_E=0.840\,87(39)\;{\rm fm}$ from \cite{Science} (which we use here as the reference value $R_E^{(0)}$) and the result $R_E=0.840\,22(56)\;{\rm fm}$ corrected for the self-consistent evaluation of the Friar term \cite{efit}. Note that in both papers on the self-consistent reevaluation (\cite{efit} and this paper) only a correction was found, while all the other theory remains the same as used in \cite{Science}.

Various extractions of the proton charge radius are summarized in Table~\ref{t:res}. The main effect of the self-consistent evaluation comes from the correction for the Friar term. The results of \cite{efit} and this paper, while correcting the Friar term, have a marginal difference which confirms that the assumption, that while calculating the $G^2$ term of $I_>$ a small area of the data around the low limit dominates, is valid. The result of the complete evaluation is marginally different from the result of the re-evaluation of the Friar term only. The strong cancellation between corrections from various contributions is rather accidental, while the marginal shift of the uncertainty is rather expected, because the corrections to $I_{\rm eTPE}$ are small and less sensitive.

\begin{table}[htbp]
\begin{center}
\begin{tabular}{c|c|c}
\hline
$R_E$ & $I_{\rm eTPE}$  & Comments \\[0.8ex]
 [fm] & [GeV$^{-3}$] &  \\[0.8ex]
\hline
0.840\,87(39) & $23.0(1.6)$ &$R_E$ from \cite{Science}, $I_{\rm eTPE}$ from \cite{birse}\\[0.8ex]
0.840\,22(56) &17.7(3.4)&$I_{\rm Fr}$ from \cite{efit}, $I_{\rm rec}$ from \cite{birse}\\[0.8ex]
0.840\,21(53)&17.7(3.3) &$I_{\rm Fr}$ from this work, $I_{\rm rec}$ from \cite{birse}\\[0.8ex]
0.840\,22(53)&17.7(3.3)&$I_{\rm eTPE}$ from this work\\[0.8ex]
\hline
\end{tabular}
\caption{The results of various evaluations of the proton charge radius. The theory follows \cite{Science} except of the treatment of the eTPE term. The eTPE term in \cite{Science} was found following \cite{birse}. The result from \cite{efit} dealt with the correction of the Friar term only. The recoil part of the eTPE is added to the $Fr$ contributions accordingly to \cite{carlson,birse}.  One of the result of {\em this paper\/} was obtained also by correcting the $Fr$ term only, while the other is a result of the complete results of the self-consistent evaluation here.\label{t:res}}
\end{center}
\end{table}

The uncertainty of the result (\ref{eq:re:result}) comes from our knowledge of the curvature of $G_E(q^2)$ at low $q$ and on our accuracy of the data on $G_E$ around $q=0.1-0.2\;{\rm GeV}$. Any improvement of those will immediately improve accuracy of determination of the proton charge radius from muonic hydrogen.

\section*{Acknowledgments}

The authors are grateful to Jose Manuel Alarc\'on, Igor Anikin,
John Arrington, Jan Bernauer, Michael Birse, Vladimir Braun, Carl Carlson,
Victor Chernyak, Michael Distler, Dieter Drechsel, Simon Eidelman,
Ron Gilman, Misha Gorshteyn, Richard Hill, Vadim Lensky, Ina Lorenz,
Judith McGovern, Ulf Meissner, Vladimir Pascalutsa, Gil Paz, Randolf Pohl,
Guy Ron, Akaki Rusetsky, Ingo Sick, and Marc Vanderhaeghen for stimulating
and useful discussions. The work was supported in part by DFG
(under grant \# HA 1457/9-1) and by RFBR (under grant \# 14-02-91332).

\appendix

\section{Fits for the electric and magnetic form factors of the proton
applied in the paper \label{app:efit}\label{s:efit}\label{app:mfit}\label{s:mfit}}

The results in the paper have been obtained by evaluation with the fits from \cite{va2011}. However, it is important to be sure that the result does not depend on details of the fits and in particular on details of their expansion around $q=0$ and their asymptotics at high $q$. Various fits for $G_E$ and $G_M$ are applied in this paper for test purposes.

The fits include two pairs of the chain-fraction fits.
Both are from Arrington and Sick, 2007, \cite{as2007}.
The first one is completely based on \cite{as2007}
\begin{eqnarray}\label{fit:as2007}
G_E(q^2)&=& \frac{1}{1 + \frac{3.44 Q^2}{1 - \frac{0.178 Q^2}{ 1 -
\frac{1.212 Q^2}{1 +
\frac{1.176Q^2}{1 - 0.284Q^2 } } } } }
\end{eqnarray}
and
\begin{eqnarray}\label{fit:mas2007}
\frac{G_M(Q^2)}{\mu_p}&=& \frac{1}{1 + \frac{3.173
        Q^2}{1 - \frac{0.314 Q^2}{ 1 - \frac{1.165Q^2}{1 + 5.619\frac{Q^2}{1 -
        1.087Q^2}}}}}\;,
\end{eqnarray}
where, $Q$ is the numerical
value for the momentum transfer $q$ in GeV.

The other pair is obtained in \cite{as2007} by applying
the two-photon correction according to \cite{bm2005}
\begin{eqnarray}\label{fit:bm2005}
G_E(q^2)&=& \frac{1}{1 + \frac{3.478Q^2}{
        1 - \frac{0.140 Q^2 }{1 - \frac{1.311Q^2}{1 + \frac{1.128Q^2}{1 - 0.233Q^2}}}}}
\end{eqnarray}
and
\begin{eqnarray}\label{fit:mbm2005}
\frac{G_M(q^2)}{\mu_p}&=& \frac{1}{1 + \frac{3.224
        Q^2}{1 - \frac{0.313 Q^2 }{ 1 - \frac{0.868 Q^2}{1 + \frac{4.278Q^2}{1 -
        1.102Q^2}}}}}\;.
\end{eqnarray}

Other fits are with the Pad\'e approximation in $q^2$,
originally introduced by Kelly, 2004, such as
\cite{kelly}
\begin{eqnarray}\label{fit:kelly}
G_E&=&\frac{1 - 0.24 \tau_p  }{1 + 10.98 \tau_p  + 12.82 \tau_p ^2 +
      0.863 \tau_p ^3}
\end{eqnarray}
and
\begin{eqnarray}\label{fit:mkelly}
\frac{G_M(q^2)}{\mu_p}&=&\frac{1 + 0.12 \tau_p  }{1 + 10.97 \tau_p  +
18.86 \tau_p ^2 +
      6.55 \tau_p ^3}\;.
\end{eqnarray}

The Pad\'e approximations have been later developed by Arrington %and Melnitchouk
et al., 2007, \cite{am2007}
\onecolumngrid
%~\\~\\~\\~\\
\begin{equation}\label{fit:am2007}
G_E-\frac{1 + 3.439 \tau_p  - 1.602\tau_p ^2 + 0.068 \tau_p ^3}{1 + 15.055 \tau_p  + 48.061 \tau_p ^2 +
      99.304 \tau_p ^3+ 0.012  \tau_p ^4 + 8.650\tau_p  ^5}
\end{equation}
\twocolumngrid
\noindent
and
%\onecolumngrid
\begin{eqnarray}\label{fit:mam2007}
\frac{G_M(q^2)}{\mu_p}&=& \frac{1 - 1.465 \tau_p  + 1.260 \tau_p ^2 +
0.262 \tau_p ^3}{1 + 9.627 \tau_p  + 11.179 \tau_p ^4 +
      13.245 \tau_p  ^5}\;.
\end{eqnarray}
%~\\
%\twocolumngrid

In paper \cite{ab2009} by Alberico %and Bilenky
et al., 2009, one electric fit is presented
\begin{equation}\label{fit:ab2009}
G_E(q^2)= \frac{1 - 0.19\tau_p }{1 + 11.12 \tau_p  + 15.16\tau_p ^2 + 21.25
\tau_p ^3}
\end{equation}
and two magnetic ones
\begin{eqnarray}\label{fit:mab20091}
\frac{G_M(q^2)}{\mu_p}&=& \frac{1 + 1.53\tau_p }{1 + 12.87 \tau_p  +
29.16\tau_p ^2 + 41.40
\tau_p ^3}\\
\frac{G_M(q^2)}{\mu_p}&=&\frac{1 + 1.09\tau_p }{1 + 12.31 \tau_p  +
25.57\tau_p ^2 + 30.61 \tau_p ^3}\label{fit:mab20092}\;.
\end{eqnarray}

The most recent Pad\'e-approximation fits (see (\ref{fit:va2011}) and (\ref{fit:mva2011})) are presented by
and Venkat et al., 2011, \cite{va2011} and have been used in the main body of the paper for all the evaluations.

One more pair of the fits with is with the inverse polynomials in $q$ from Bosted, 1995, \cite{bo1994}
\begin{equation}\label{fit:bo1994}
G_E(q^2)= \frac{1}{1 + 0.62Q + 0.68 Q^2 + 2.8 Q^3 + 0.83Q^4}
\end{equation}
and
\onecolumngrid
%~\\~\\
\begin{equation}\label{fit:mbo1994}
\frac{G_M(q^2)}{\mu_p}= \left(1 + 0.35Q + 2.44 Q^2 + 0.50 Q^3+
1.04Q^4 + 0.34Q^5\right)^{-1}\;.
\end{equation}
%~\\
\twocolumngrid
\noindent
That is a
phenomenological fit designed to be used for medium and high $q$. It
is not expected to be appropriate at low $q$. Providing a reasonably
good approximation at medium momentum transfer, the fit apparently
has incorrect low-$q$ behavior and incorrect analytic properties
such as a branch point at $q^2=0$.

\end{document}